\documentclass[useAMS,usenatbib,usedcolumn]{mn2e}
\usepackage{epsfig}
\usepackage{amsmath}

\newcommand{\be}{\begin{equation}}
\newcommand{\beq}{\begin{equation}}
\newcommand{\ba}{\begin{eqnarray}}
\newcommand{\ee}{\end{equation}}
\newcommand{\eeq}{\end{equation}}
\newcommand{\ea}{\end{eqnarray}}

\def\lsim{~\rlap{$<$}{\lower 1.0ex\hbox{$\sim$}}}

\def\gsim{~\rlap{$>$}{\lower 1.0ex\hbox{$\sim$}}}

\title[NIR type Ia SNe]{Near-infrared observations of type Ia supernovae: The best known standard candle for cosmology}

\author[Barone-Nugent et al.]{R. L. Barone-Nugent$^{1}$, C. Lidman$^{2, 3}$, J. S. B. Wyithe$^{1}$, J. Mould$^{4}$, D. A. Howell$^{5, 6}$, \newauthor I. M. Hook$^{7, 8}$, M. Sullivan$^{7}$, P. E. Nugent$^{9}$, I. Arcavi$^{10}$, S. B. Cenko$^{11}$, J. Cooke$^{4}$, \newauthor A. Gal-Yam$^{10}$, E. Y. Hsiao$^{12}$, M. M. Kasliwal$^{12}$, K. Maguire$^{7}$, E. Ofek$^{10}$, \newauthor D. Poznanski$^{13}$, D. Xu$^{10}$\\
$^{1}$ School of Physics, University of Melbourne, Parkville, Victoria, Australia\\
$^{2}$ Australian Astronomical Observatory, Epping, New South Wales, Australia\\
$^{3}$ ARC Centre of Excellence for All-sky Astrophysics (CAASTRO)\\
$^{4}$ Swinburne University, Hawthorn, Victoria, Australia\\
$^{5}$ Las Cumbres Observatory Global Telescope Network, 6740 Cortona Dr., Suite 102, Goleta, CA 93117, USA\\
$^{6}$ Department of Physics, University of California, Santa Barbara, Santa Barbara, CA 93106\\
$^{7}$ Department of Physics (Astrophysics), University of Oxford, DWB, Keble Road, Oxford OX1 3RH, UK\\
$^{8}$ INAF, Osservatorio Astronomico di Roma, via Frascati 33, 00040 Monteporzio (RM), Italy\\
$^{9}$ E.O. Lawrence Berkeley National Lab, 1 Cyclotron Rd., Berkeley, CA 94720\\
$^{10}$ Weizmann Institute of Science, Rehovot 76100, Israel\\
$^{11}$ Department of Astronomy, University of California, Berkeley, CA 94720-3411, USA\\
$^{12}$ Observatories of the Carnegie Institution for Science, 813 Santa Barbara St, Pasadena CA 91101 USA\\
$^{13}$ School of Physics and Astronomy, Tel Aviv University, Tel Aviv 69978, Israel\\
Email: robertbn@student.unimelb.edu.au; clidman@aao.gov.au; swyithe@unimelb.edu.au; jmould@astro.swin.edu.au,\\ sullivan@astro.ox.ac.uk}
\begin{document}
\date{\today}
\maketitle
\label{firstpage}
\begin{abstract}
\noindent We present an analysis of the Hubble diagram for 12 normal type Ia supernovae (SNe Ia) observed in the near-infrared $J$ and $H$ bands. We select SNe exclusively from the redshift range $0.03 < z <0.09$ to reduce uncertainties coming from peculiar velocities while remaining in a cosmologically well-understood region. All of the SNe in our sample exhibit no spectral or \textit{B}-band light-curve peculiarities and lie in the \textit{B}-band stretch range of $0.8$ - $1.15$. Our results suggest that SNe Ia observed in the near-infrared (NIR) are the best known standard candles. We fit previously determined NIR light-curve templates to new high-precision data to derive peak magnitudes and to determine the scatter about the Hubble line. Photometry of the 12 SNe is presented in the natural system. Using a standard cosmology of $(H_{0},\Omega_{m},\Omega_{\Lambda}) = (70,0.27,0.73)$ we find a median $J$-band absolute magnitude of $M_{\textnormal{J}} = -18.39$ with a scatter of $\sigma_{\textnormal{J}} =0.116$ and a median $H$-band absolute magnitude of $M_{\textnormal{H}} = -18.36$ with a scatter of $\sigma_{\textnormal{H}} =0.085$. The scatter in the $H$ band is the smallest yet measured. We search for correlations between residuals in the $J$- and $H$-band Hubble diagrams and SN properties, such as SN colour, B-band stretch and the projected distance from host-galaxy centre. The only significant correlation is between the $J$-band Hubble residual and the $J$-$H$ pseudo-colour. We also examine how the scatter changes when fewer points in the near-infrared are used to constrain the light curve. With a single point in the $H$ band taken anywhere from 10 days before to 15 days after B-band maximum light and a prior on the date of $H$-band maximum set from the date of B-band maximum, we find that we can measure distances to an accuracy of 6\%. The precision of SNe Ia in the NIR provides new opportunities for precision measurements of both the expansion history of the universe and peculiar velocities of nearby galaxies.\end{abstract}
\section{Introduction}
SNe Ia are widely accepted to be excellent standardizable candles at optical wavelengths. The development of various empirical corrections, which can relate peak luminosity and light-curve shape \citep{phillips1993absolute, hamuy1996bvri}, SN colour \citep{tripp1998two, jha2007improved}, spectral information \citep{bronder2007snls, walker2011supernova, silver2012berkeley} and host-galaxy mass \citep{sullivan2010dependence}, allow SNe Ia peak magnitudes to serve as distance indicators with a corrected scatter\footnote{We use the rms about the mean as a measure of the scatter.} of as low as $0.13$ magnitudes \citep{silver2012berkeley, conley2011supernova}.\\
\indent There has been growing evidence, though, that SNe Ia may be more accurate in the near-infrared (NIR) (for a recent review, see \citealt{phillips2011near}). Near-infrared light experiences less attenuation from dust than light at optical wavelengths, and theoretical models predict a smaller intrinsic dispersion in the near-infrared peak magniude \citep{kasen2006secondary}. Furthermore, there appears to be little or no relationship between light-curve shape and peak luminosity \citep{wood2008type, folatelli2010carnegie, mandel2011type}, meaning no empirical corrections need be applied.\\ 
\indent Recent studies \citep{meikle2000absolute, wood2008type, folatelli2010carnegie, krisciunas2004hubble, kattner2012standardizability, mandel2011type} have shown that SNe Ia in the NIR can be as reliable as corrected SNe Ia observed at optical wavelengths, but these studies have had some limitations. Foremost among these limitations is that the SNe were observed at distances that are not sufficiently large to place them in the Hubble flow (defined here as $z > 0.03$) and hence are affected by peculiar velocities. Here we present a study of a sample of Type Ia SNe selected to lie exclusively in the redshift range $0.03 < z <0.09$ in order to minimise redshift uncertainty due to peculiar velocity. We use NIR images on 8m class telescopes to observe supernovae in our sample, resulting in photometry that is more precise than that achieved for the handful of supernovae that have been observed beyond $z = 0.03$.\\

\vspace{-7mm}
\section{SNe Ia sample}
\indent The SNe Ia used in our sample are summarised in Table \ref{Table:resultsH&J}. These SNe Ia were discovered by the Palomar Transient Factory (PTF; \citealt{rau2009exploring, law2009palomar}), spectroscopically confirmed, observed in the $g,r,i$ bands (Maguire et al. in prep.) at the Liverpool Telescope and followed up in both the NIR $J$ and $H$ bands, each with four epochs of observation. PTF09dlc was observed using HAWK-I \citep{casali2006hawk} on ESO's 8.1m VLT, and all others were observed using NIRI\footnote{http://www.eso.org/sci/facilities/paranal/instruments/hawki/ and http://www.gemini.edu/sciops/instruments/niri/}, the NIR Imager and Spectrometer \citep{hodapp2000gemini} on Gemini Observatory's 8.2m Gemini North. Reference images were taken $\sim 1$ year after initial observations so that host-galaxy light could be removed.\\ 
\indent The data were processed in a standard manner using IRAF\footnote{IRAF is distributed by the National Optical Astronomy Observatories which are operated by the Association of Universities for Research in Astronomy, Inc. under the cooperative agreement with the National Science Foundation} and our own scripts. Darks were used to remove the pedestal, and flats were used to remove pixel-to-pixel sensitivity variations. We used the {\tt XDIMSUM} package in IRAF to remove the sky and a modified version of ISIS2 \citep{alard2000image} to subtract the reference image of each SN from the images with SN light, leaving just light of the SN. Zero-points were derived from standards found in the \citet{persson1998new} catalogue. All magnitudes are reported in the natural system of each instrument. K-corrections were calculated using the revised spectral template of \citet{hsiao2007k} and corrections for galactic dust extinction were applied. We did not warp the spectral template to match the observed colour. \\
\indent For the SNe PTF09dlc, PTF10hmv, PTF10mwb, PTF10nlg, PTF10tce, PTF10ufj, PTF10wnm and PTF10xyt, the first observing epoch was before NIR maximum. For the remaining SNe, the first observing epoch was near NIR maximum. When fitting light-curves for the SNe, we included the date of $B$-band maximum as an extra prior. The $B$-band maximum was determined using SiFTO \citep{conley2008sifto}, as was the $B$-band stretch (Maguire et al. in prep.).\\
\indent The $B$-band stretches range from 0.8 to 1.15, so there are no fast decliners in our sample. Nor are there any heavily reddened supernovae.  Furthermore, all of the SNe appear to be spectroscopically normal. The absence of fast decliners and heavily reddened SNe is a potential limitation of the current sample.\\
\indent The photometry for all SNe is presented in Table \ref{table:photometry}.
\begin{table*}
\begin{center}
\begin{tabular}{l | c c c}
\hline
\hline
SN name & MJD & m$_{J}$ & m$_{H}$ \\ 
\hline
PTF09dlc          &  55068.14  & 19.803 $\pm$  0.023  & 18.813 $\pm$  0.021 \\
                        &  55073.11  & 19.906 $\pm$  0.040  & 18.933 $\pm$  0.044 \\
                        &  55085.16  & 20.997 $\pm$  0.059  & 19.646 $\pm$  0.050 \\
                        &  55093.18  & 21.399 $\pm$  0.067  & 19.635 $\pm$  0.028\\
PTF10hdv         &  55340.25  & 19.084 $\pm$  0.044  & 18.516 $\pm$  0.052\\
                        &  55346.25  & 19.407 $\pm$  0.057    & 18.766 $\pm$  0.052\\
                        &  55350.25  & 19.787 $\pm$  0.085  & 18.920 $\pm$  0.065\\
                        &  55354.41  & 20.266 $\pm$  0.139  & 19.048 $\pm$  0.085\\
PTF10hmv         &  55341.26  & - & 17.887 $\pm$  0.019 \\
                        &  55346.28  & -  & 17.501 $\pm$  0.026 \\
                        &  55351.30  & -  & 17.595 $\pm$  0.017\\
                        &  55357.25  & -  & 17.823 $\pm$  0.054\\
                        &  55358.25  & -  & 17.825 $\pm$  0.019\\
                        &  55359.26  & -  & 17.826 $\pm$  0.018 \\
PTF10mwb        &  55378.41  & 19.066 $\pm$  0.031  & 18.429 $\pm$  0.030\\
                        &  55381.43  & 18.677 $\pm$  0.027 & 17.765 $\pm$  0.014\\
                        &  55386.41  & 17.683 $\pm$  0.026  & 17.484 $\pm$  0.012\\
                        &  55391.43  & 18.202 $\pm$  0.022  & 17.569 $\pm$  0.013\\
                        &  55396.39  & 18.584 $\pm$  0.023  & 17.760 $\pm$  0.014\\
PTF10ndc         &  55386.43  & 19.898 $\pm$  0.062  & 19.249 $\pm$  0.079\\
                        &  55389.38  & 20.110 $\pm$  0.088  & 19.468 $\pm$  0.099\\
                        &  55393.39  & 20.333 $\pm$  0.092   & 19.488 $\pm$  0.107\\
                        &  55395.35  & 20.496 $\pm$  0.101           & 19.519 $\pm$  0.118\\
PTF10nlg         &  55386.39  & 19.352 $\pm$  0.044  & 18.694 $\pm$  0.039\\
                        &  55387.33  & -  & 18.500 $\pm$  0.041\\
                        &  55389.35  & -  & 18.571 $\pm$  0.035\\
                        &  55393.36  & 19.679 $\pm$  0.073   & 18.770 $\pm$  0.044\\
                        &  55395.37  & 19.816 $\pm$  0.059           & 18.791 $\pm$  0.057\\
PTF10qyx        &  55422.63  & 19.652 $\pm$  0.069 & 19.005 $\pm$  0.046\\
                        &  55424.56  & 19.701 $\pm$  0.045  & 19.121 $\pm$  0.049\\
                        &  55428.64  & 20.156 $\pm$  0.076   & 19.487 $\pm$  0.102\\
                        &  55433.57  & 20.654 $\pm$  0.104                   & 19.736 $\pm$  0.088\\
PTF10tce         &  55435.41  & 18.634 $\pm$  0.040  & 18.047 $\pm$  0.032\\
                        &  55438.34  & 18.592 $\pm$  0.036   & 17.943 $\pm$  0.026\\
                        &  55443.50  & 18.778 $\pm$  0.041   & 18.164 $\pm$  0.028\\
                        &  55448.53  & 19.175 $\pm$  0.059                    & 18.352 $\pm$  0.035\\
PTF10ufj          &  55447.55  & 20.496 $\pm$  0.112   & 19.444 $\pm$  0.079\\
                        &  55450.64  & 20.387 $\pm$  0.093   & 19.241 $\pm$  0.068\\
                        &  55455.50  & 20.312 $\pm$  0.100   & 19.202 $\pm$  0.061\\
                        &  55460.58  & 20.996 $\pm$  0.238                  & 19.544 $\pm$  0.097\\
PTF10wnm       &  55469.38  & 19.513 $\pm$  0.055   & 18.952 $\pm$  0.044 \\
                        &  55471.48  & 19.443 $\pm$  0.052   & 18.822 $\pm$  0.040\\
                        &  55476.27  & 19.482 $\pm$  0.053   & 18.984 $\pm$  0.069\\
                        &  55481.28  & 19.883 $\pm$  0.072        & 19.262 $\pm$  0.126\\
PTF10wof         &  55470.33  & 19.086 $\pm$  0.030   & 18.488 $\pm$  0.033\\
                        &  55473.29  & 19.201 $\pm$  0.039  & 18.592 $\pm$  0.042\\
                        &  55478.24  & 19.491 $\pm$  0.063  & 18.855 $\pm$  0.051 \\
                        &  55483.29  & 20.071 $\pm$  0.071          & 19.173 $\pm$  0.061\\
PTF10xyt         &  55486.26  & 19.116 $\pm$  0.051  & 19.071 $\pm$  0.118\\
                        &  55492.24  & 19.277 $\pm$  0.047  & 19.438 $\pm$  0.179\\
                        &  55496.21  & -  & 19.735 $\pm$  0.238\\
                        &  55498.20  & -  & 19.596 $\pm$  0.227\\
\hline
\hline
\end{tabular}
\caption{The $J$- and $H$-band photometry for each epoch of each SNe in our sample. Apparent magnitudes have been corrected for Galactic dust extinction but not K-corrected.}
\label{table:photometry}
\end{center}
\end{table*}

\vspace{-7mm}
\section{Template Fitting}
\begin{figure*}
\begin{minipage}[b]{1\linewidth}
\centering
\includegraphics[scale=0.27]{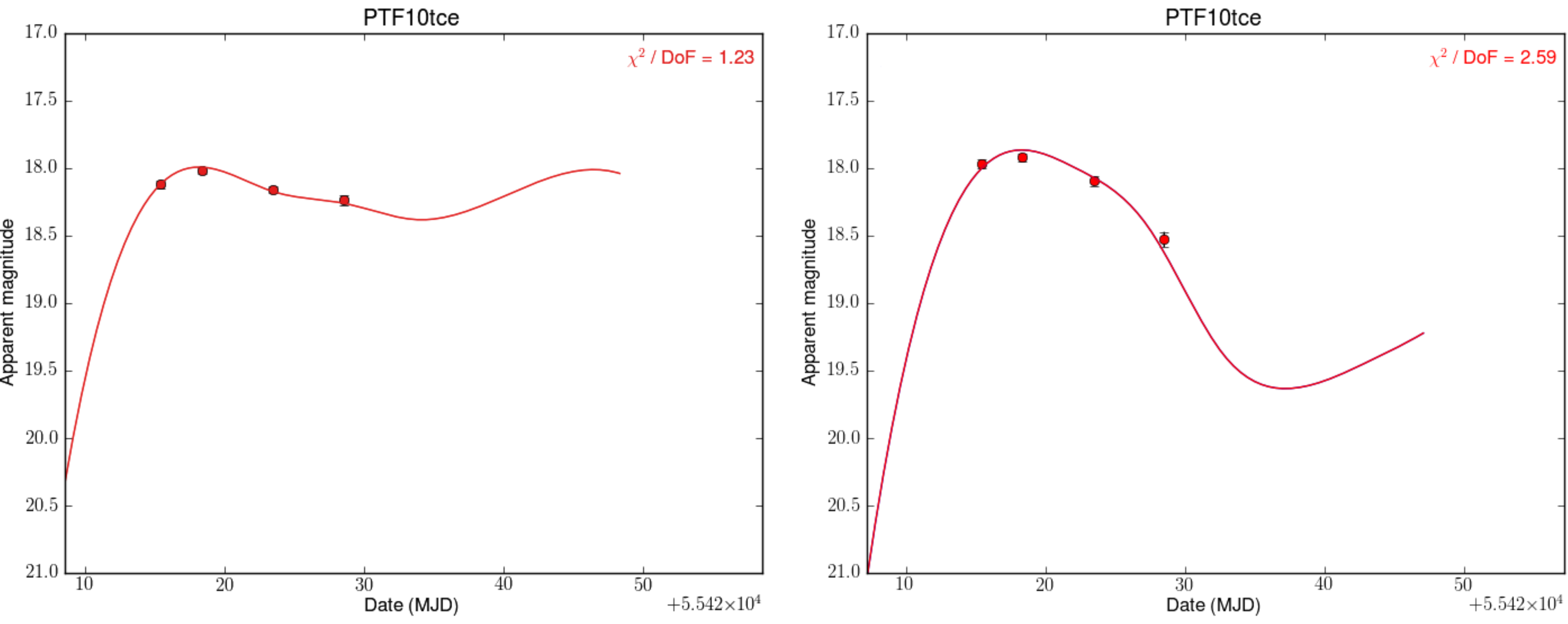}
\caption{The FLIRT template fitted to the light-curves for PTF10tce in the $H$ band \textit{\textbf{(left)}} and the $J$band \textit{\textbf{(right)}}.}
\vspace{-2mm}
\label{fig:lc}
\end{minipage}
\end{figure*}

\indent To find the light-curve maxima, we fit the Flexible Light-curve InfraRed Template (FLIRT) presented in \citet{mandel2009type} to our K-corrected photometry\footnote{Photometry to be published in Barone-Nugent et al. (in prep.).}. As we did not observe all SNe Ia before maximum light, we fit the template to the data with an additional constraint. Concurrent observations of SNe Ia at NIR and optical wavelengths provide evidence that NIR maxima occur $\sim 5$ days before $B$-band maximum \citep{meikle2000absolute, krisciunas2004hubble}. We use this prior assumption as a constraint on the date of NIR maximum, $t_{p}$, with a $1-$$\sigma$ error of $\sigma_{p} =$ 1 day, imposed on the likelihood function. By fitting the light-curve template to the SNe Ia observed before maximum without a prior, we determine that the $J$-band maximum occurs $5.36$ days before the $B$-band maximum with a standard deviation of $0.74$ days and the $H$-band maximum occurs $4.28$ days before with a standard deviation of $0.70$ days. Figure \ref{fig:lc} displays the template-fitted light-curve for PTF10tce.\\
\indent The errors in the peak apparent magnitudes in both the $J$ and $H$ band were calculated via 100 Monte Carlo simulations of template fitting to the data, including repeated Gaussian-distributed random samplings within the errors of each point.\\
\indent To convert apparent magnitudes to absolute magnitudes, we determine the distance modulus for each SN, calculated using the standard flat cosmology of $(\textnormal{H}_{0},\Omega_{m},\Omega_{\Lambda}) = (70,0.27,0.73)$. The redshift measurement for PTF10ufj is uncertain since the redshift is measured from the supernova and not the host. In this case, we have assumed an uncertainty of $\sigma_{z} = 0.005$ (Maguire et al. in prep.). For all other redshifts, the errors make negligible difference to the distance modulus.\\

\vspace{-7mm}
\section{Results}
\begin{figure*}
\begin{minipage}[b]{1\linewidth}
\centering
\includegraphics[scale=0.24]{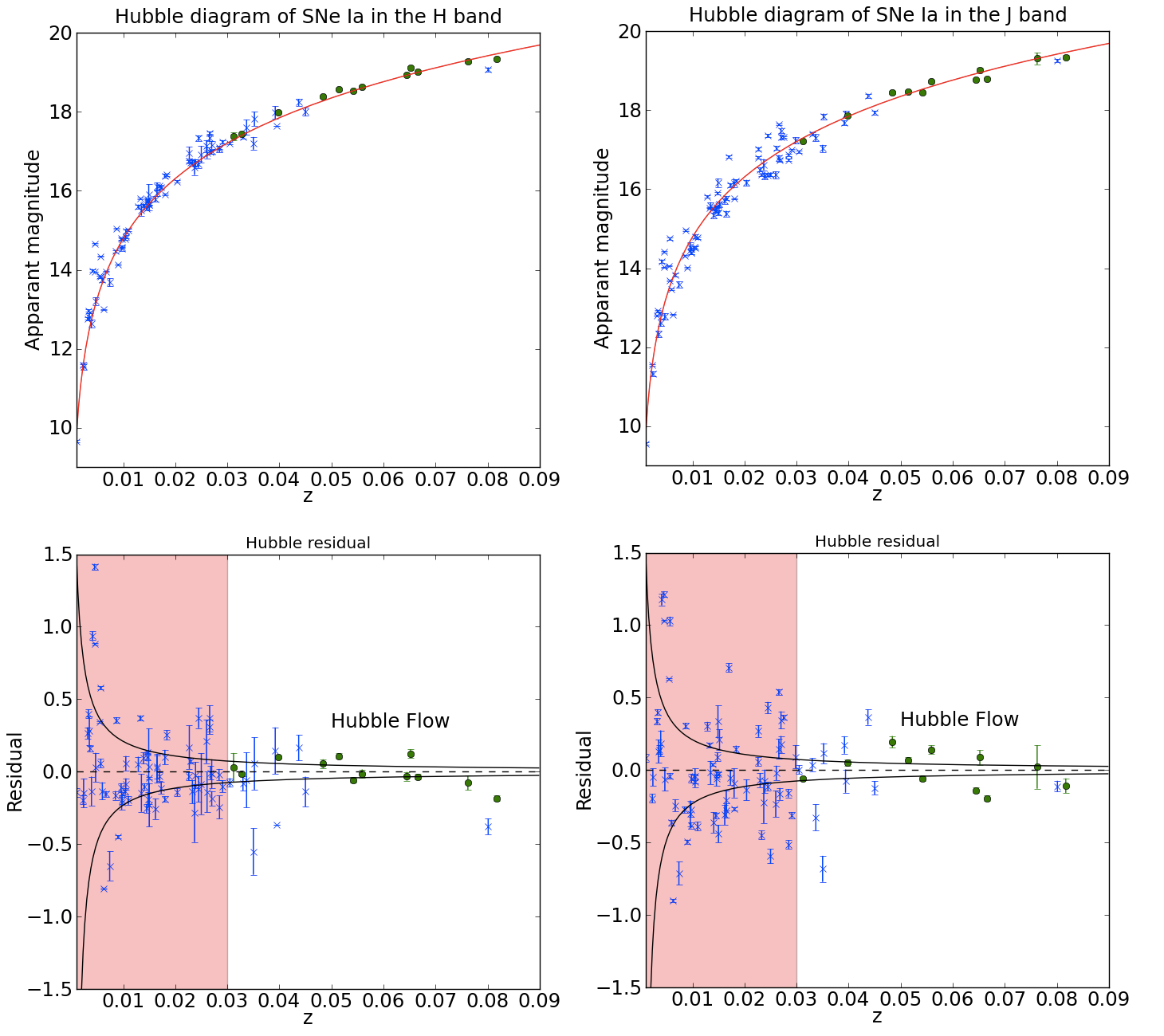}
\caption{The upper panels shows the Hubble diagrams of our sample of 12 SNe (green circles) in the \textit{H} band \textit{\textbf{(left)}} and the \textit{J} band \textit{\textbf{(right)}}, including the \citet{wood2008type}, \citet{folatelli2010carnegie} and \citet{krisciunas2004hubble} samples (blue crosses) for comparison. The red line represents the apparent magnitude that would be observed assuming a constant absolute magnitude of $-18.36$ (\textit{H} band) and $-18.39$ (\textit{J} band). The lower panels show the Hubble residual, i.e. the deviation from the red line. The shaded red region, $z < 0.03$, is the region excluded in our sample due to the associated peculiar velocity errors. The solid black lines represent the change in the distance modulus due to a peculiar velocity of $\pm300$ km\,s$^{-1}$}
\label{fig:hubble_H}
\end{minipage}
\end{figure*}

\indent The peak apparent magnitudes for each SN Ia in the \textit{J} and \textit{H} band are quoted in Table \ref{Table:resultsH&J}. In the $H$ band, the median absolute magnitude of the supernovae in our sample is $M_{\textnormal{H}}=-18.36 \pm 0.04$. This compares reasonably well with the median magnitudes of supernovae in \cite{folatelli2010carnegie}, \citet{wood2008type} and \citet{krisciunas2004hubble} for which we derive $M_{\textnormal{H}}=-18.37 \pm 0.07$, $M_{\textnormal{H}}=-18.29 \pm 0.04$, and $M_{\textnormal{H}}=-18.43 \pm 0.06$, respectively. The errors on the median are computed by jackknife resampling. In part, the difference in magnitudes comes from the limited number of supernovae in each of the samples. The differences may also be due to the differences in the photometric systems that each author adopts. These differences are generally smaller than 0.04 mag.\\
\indent In the $J$ band, the variations between samples are larger and more difficult to understand. We find a median magnitude of  $M_{\textnormal{J}}=-18.39 \pm 0.06$, compared to median magnitudes of $M_{\textnormal{J}}=-18.48 \pm 0.05$, $M_{\textnormal{J}}=-18.29 \pm 0.10$, and $M_{\textnormal{J}}=-18.73 \pm 0.05$ for the supernovae in \cite{folatelli2010carnegie}, \citet{wood2008type} and \citet{krisciunas2004hubble}. The errors were calculated as for the $H$ band. The largest source of uncertainty is due to sample size. Difference due to photometric systems are generally smaller. In order to constrain the dark energy equation of state beyond limits currently derived using supernovae observed in the optical, the mean magnitude needs to be determined to a level of accuracy that is better than 0.01 mag.\\
\indent The $J$- and $H$-band Hubble diagrams are shown in Figure \ref{fig:hubble_H}. Our results are compared with the surveys of \citet{wood2008type} and \citet{folatelli2010carnegie}. Our observed scatter about the mean absolute magnitude of $\sigma_{\textnormal{H}} = 0.085 \pm 0.016$ mag and $\sigma_{J} = 0.116 \pm 0.027$ mag represent a considerable improvement on previous samples, including \citet{wood2008type} who observed $\sigma_{\textnormal{H}} = 0.28$ mag and $\sigma_{J} = 0.41$ mag, \citet{krisciunas2004hubble} who observed $\sigma_{\textnormal{H}} = 0.17$ mag and $\sigma_{J} = 0.13$ mag, \citet{folatelli2010carnegie} who observed $\sigma_{\textnormal{H}} = 0.33$ mag and $\sigma_{J} = 0.37$ mag and \citet{mandel2009type} who observed $\sigma_{\textnormal{H}} = 0.11$ mag and $\sigma_{J} = 0.17$ mag. The values for $\sigma_{\textnormal{J}}$ and $\sigma_{\textnormal{H}}$ presented here are the scatter in the magnitude at \textit{J}-max and \textit{H}-max, respectively.\\
\indent Photometric errors make little contribution to the scatter we measure. Peculiar velocities only make a significant contribution if they are large (i.e. $\sim 300$km\,s$^{-1}$). The intrinsic scatters we measure, assuming peculiar velocities of $300$ km\,s$^{-1}$ are $\sigma_{\textnormal{int}} = 0.101$ in the $J$ band and $\sigma_{\textnormal{int}} = 0.065$ in the $H$ band. For peculiar velocities of $150$ km\,s$^{-1}$ we obtain intrinsic scatters of $\sigma_{\textnormal{int}} = 0.109$ in the $J$ band and $\sigma_{\textnormal{int}} = 0.077$ in the $H$ band. Part of the scatter we observe may also be due to spectral variability affecting the K-corrections.\\
\indent We note that the light-curves of some of the SN in our sample are not adequately described by the light-curve template that we have used. This is because there is a diversity in light-curve shapes, which may lead to opportunities to reduce the scatter about the Hubble line further. We investigated this by recomputing the scatter after removing SNe with poor light-curve fits. If we eliminate light-curves with a $\chi^{2}/$Degrees of Freedom greater than $5$ (a chance probability of $Q = 0.002$), we reject PTF09dlc and PTF10mwb in the $J$ band and PF09dlc, PTF10hmv and PTF10mwb in the $H$ band. The resulting values of scatter in the $J$ and $H$ bands are $0.105$ and $0.097$ respectively. The scatter in the $H$ band increases marginally but is noticeably lower for the $J$ band. PTF09dlc, PTF10hmv and PTF10mwb exhibit no spectral or $B$-band light-curve peculiarities. However, applying such cuts for the purpose of reducing the scatter needs to be done with care since they may introduce biases that may affect the accuracy with which cosmological parameters can be measured.\\
\begin{table*}
\vspace{-0mm}
\begin{center}
\begin{tabular}{l | l l l l l l}
\hline \hline
Supernova & RA(J2000) & DEC (J2000) & $z_{helio}$ & $z_{CMB}$ & $m_{\textnormal{H,max}}$ & $m_{\textnormal{J,max}}$\\
\hline
PTF09dlc 		& 21:46:35.5 & +06:23:23.5 & $0.0678$ & $0.0666$ &  $19.010^{+0.016}_{-0.018}$		& $18.782^{+0.015}_{-0.020}$\\
PTF10hdv 	& 12:07:43.4 & +41:29:27.9 & $0.0534$ & $0.0542$ &  $18.521^{+0.031}_{-0.037}$ 	& $18.453^{+0.026}_{-0.026}$ \\
PTF10hmv 	& 12:11:33.0 & +47:16:29.8 & $0.0320$ & $0.0327$ &  $17.431^{+0.009}_{-0.008}$	& -\\
PTF10mwb 	& 17:17:50.0 & +40:52:52.1 & $0.0313$ & $0.0312$ &  $17.371^{+0.007}_{-0.007}$		& $17.213^{+0.010}_{-0.009}$ \\
PTF10ndc 	& 17:19:50.2 & +28:31:57.5 & $0.0818$ & $0.0817$ &  $19.330^{+0.085}_{-0.133}$		& $19.337^{+0.051}_{-0.043}$	 \\
PTF10nlg 		& 16:50:34.5 & +60:16:35.0 & $0.0560$ & $0.0559$ &  $18.634^{+0.018}_{-0.020}$		& $18.720^{+0.032}_{-0.029}$	\\
PTF10qyx 	& 02:27:12.1 & -04:31:04.8 & $0.0660$ & $0.0652$ &  $19.122^{+0.031}_{-0.027}$		& $19.016^{+0.042}_{-0.053}$\\
PTF10tce 		& 23:19:11.0 & +09:11:54.2 & $0.0410$ & $0.0398$  & $17.988^{+0.016}_{-0.011}$		& $17.868^{+0.018}_{-0.021}$	\\
PTF10ufj 		& 02:25:39.1 & +24:45:53.2 & $0.077 \pm 0.005$ & $0.0762 \pm 0.005$ & $19.276^{+0.036}_{-0.047}$ 	& $19.306^{+0.063}_{-0.051}$\\
PTF10wnm 	& 0:13:47.26 & +27:02:26.0 & $0.0656$ & $0.0645$ & $18.941^{+0.028}_{-0.037}$	& $18.764^{+0.022}_{-0.021}$\\
PTF10wof 	& 23:32:41.8 & +15:21:31.7 & $0.0526$ & $0.0514$ & $18.567^{+0.025}_{-0.024}$	& $18.458^{+0.022}_{-0.021}$\\
PTF10xyt		& 23:19:02.4 & +13:47:26.8 & $0.0496$ & $0.0484$ &  $18.379^{+0.026}_{-0.023}$	& $18.450^{+0.036}_{-0.037}$\\
\hline \hline
\end{tabular}
\caption{The sample of SNe Ia showing their redshifts and $J$- and $H$-band peak apparent magnitudes. PTF10ndc, PTF10tce, PTF10ufj, PTF10wof, PTF10xyt were located by the citizen science Galaxy Zoo: Supernovae project \citep{smith2011galaxy}.}
\label{Table:resultsH&J}
\end{center}
\vspace{-4mm}
\end{table*} 
\indent The relationship between the peak magnitude in both filters and pseudo-colour ($m_{\textnormal{J,max}}$-$m_{\textnormal{H,max}}$) was also considered. In the optical, there is a strong correlation between colour and luminosity \citep{tripp1998two, jha2007improved}. However, we find no obvious correlation between the $H$-band residual and J$_{\textnormal{max}}$-H$_{\textnormal{max}}$ pseudo-colour in our sample. On the other hand, we do find a correlation between the $J$-band residual and pseudo-colour. From the least-squares residuals in the $y$-direction, we find a relation for the line of best fit, where the errors on the coefficients are $1-\sigma$ dispersions calculated via 1000 Monte Carlo simulations by randomly varying the points within their error bars:
\begin{equation} \label{eq:bestfit}
\begin{split}
J\textnormal{-band residual} = (0.91 \pm 0.2) \times &(M_{\textnormal{J,max}}-M_{\textnormal{H,max}}) \\
&+ (0.05 \pm 0.02)
\end{split}
\end{equation}
with a scatter about the line of best fit of $\sigma = 0.071$. \\
\indent We also considered the relationship between the peak magnitude in both filters with the angular distance between the SN and the centre of the host galaxy and $B$-band stretch. However, we find no correlations between any of these quantities. It should be noted that the \textit{B}-band stretch range covered by this sample is limited.
\subsection{Template fitting with a subset of points}
\indent The previous discussion refers to SNe Ia light-curves consisting of 4 data points, which we used to fit the light-curve templates. In this section we discuss the optimal number of epochs that should be observed in the NIR given finite telescope time.\\
\indent We have already calculated the peak magnitude, $M_{\textnormal{J,H}}$, for each SNe Ia using all data available. We now calculate the peak magnitudes using a subset of the data points of size $x$, denoted by $M_{x-pts}$, and find the scatter of these magnitudes around $M_{\textnormal{J,H}}$. The total scatter in the sample using $x$ points is denoted by $\sigma_{\textnormal{J,H},tot}$. The results for these quantities based on our SNe Ia sample are presented in Table \ref{Table:4ptsJ}.\\
\begin{table}
\begin{center}
\begin{tabular}{ l | c l c l}
\hline \hline
$x$ & $\overline{M_{\textnormal{J}}}_{x-pts}-M_{\textnormal{J}}$ & $\sigma_{\textnormal{J}, \textnormal{tot}}$ & $\overline{M_{\textnormal{H}}}_{x-pts}-M_{\textnormal{H}}$ & $\sigma_{\textnormal{H}, \textnormal{tot}}$ \\
\hline
1 & 0.002  & 0.146 & 0.007 & 0.116\\
2 & 0.012  & 0.126 & 0.002 & 0.096\\
3 & 0.008  & 0.120 & 0.002 & 0.089\\
4 & 0.000  & 0.116 & 0.000 & 0.085\\
\hline \hline
\end{tabular}
\caption{The mean peak magnitude and dispersion in the $J$ and $H$ bands when using light-curves with $x$ points.}
\label{Table:4ptsJ}
\end{center}
\vspace{-4mm}
\end{table}
\indent We find in all cases that there is approximately one-hundredth of a magnitude difference in the mean peak magnitude or less between different numbers of epochs used. However there is additional scatter around the mean which decreases as more points are added. This illustrates the exchange between amount of telescope time used and dispersion measured. Table \ref{Table:4ptsJ} suggests that the increased accuracy from 3 points to 4 is $\sim 0.01$ mag or less in both the $J$ and $H$ bands. Furthermore, using just one point with a known date of maximum can be used while only sacrificing $\sim 0.04$ mag of accuracy (i.e. 1-$\sigma$ scatter) in both the $J$ and $H$ bands. Importantly, using just one point in the $J$ band gives a dispersion that is similar to the best standardized SNe Ia observed in the optical, and using just one point in the $H$ band can further improve the accuracy of SNe Ia observed in the optical. We also find that, within the range of epochs covered by our data (approximately $10$ days before to $15$ days after $B$-band maximum), the amount of additional scatter introduced due to using only one point has no dependence on when this point is taken with respect to $B$-band maximum. Importantly, single observations after \textit{B}-band peak are equally as effective as points prior to the peak.\\
\indent We may now compare the standard error of the mean (SEM) for each subset of points. The SEM for a sample of $n$ points is given by,
\beq
\textnormal{SEM} = \frac{\sigma}{\sqrt{n}}
\ee
where $\sigma$ is the standard deviation of the sample. As an illustrative example, we compare observing $3$ SNe over $4$ epochs, observing $4$ SNe over $3$ epochs, observing $6$ SNe over $2$ epochs and observing $12$ SNe for a single epoch. In all cases, we assume that we have well-sampled optical light-curves for each SN Ia. Each of these would require equal allocation of telescope time. The relative SEM for each method are summarised in Table \ref{Table:relSEM}. We find that observing four times the number of SNe with only a single epoch delivers a lower SEM than observing fewer SNe over more than one epoch. The larger sample size more than compensates for the higher uncertainty in the peak magnitude.
\begin{table}
\begin{center}
\begin{tabular}{l | l l l l l l}
\hline \hline
$x$ & Sample size & $\sigma_{\textnormal{H}}$ & SEM$_{\textnormal{H}}$ & $\sigma_{\textnormal{J}}$ & SEM$_{\textnormal{J}}$\\
\hline
$1$ & $12$ & $0.116$ & $0.033$ & $0.146$ & $0.042$\\
$2$ & $6$ & $0.096$ & $0.039$ & $0.126$ & $0.051$\\
$3$ & $4$ & $0.089$ & $0.045$ & $0.120$ & $0.060$\\
$4$ & $3$ & $0.085$ & $0.049$ & $0.116$ & $0.067$\\
\hline \hline
\end{tabular}
\caption{The standard error of the mean for an example of 12 observing nights distributed among $3$, $4$, $6$ and $12$ SNe.} 
\label{Table:relSEM}
\end{center}
\vspace{-4mm}
\end{table}
\vspace{-4mm}
\section{Conclusion}
\indent We have presented NIR light curves of 12 SNe Ia that are in the Hubble flow. We find that the intrinsic scatter in peak luminosities of type Ia supernovae in the NIR $J$ and $H$ bands are smaller than previously thought. This is the first sample to display an $H$-band rms scatter as small as $\sigma_{\textnormal{H}} = 0.085 \pm 0.016$  (with a median peak magnitude of $M_{\textnormal{H}} = -18.36$). Our observed $J$-band rms scatter of $\sigma_{\textnormal{J}} = 0.116 \pm 0.027$ (with a median peak magnitude of $M_{\textnormal{J}} = -18.39$) is smaller than reported elsewhere. These results provide distance errors of $\sim 4\%$ using $H$ band SNe, making them the most precise standard candles for cosmology.\\
\indent We have also shown that if concurrent optical observations are made, we may use a predicted date of NIR maximum as a constraint when fitting the light-curve. With this constraint, we may use as few as one NIR observation within $\sim 5-10$ days of NIR max per supernova while still achieving scatters of $\sigma_{\textnormal{H}} = 0.116$ and $\sigma_{J} = 0.146$. As surveys improve over the coming years and more SNe Ia are discovered, single-night near-infrared observations undertaken concurrently with optical observations will be the most efficient and accurate way to construct samples.\\

{\bf Acknowledgments} 
Based on data collected at the ESO VLT (program number 083.A-0480) and observations obtained at the Gemini Observatory (program numbers GN2010A-Q-16, GN2010B-Q-17, GN2011A-Q-11 and GN-2011B-Q-21), which is operated by the Association of Universities for Research in Astronomy, Inc., under a cooperative agreement with the NSF on behalf of the Gemini partnership: the National Science Foundation (United States), the Science and Technology Facilities Council (United Kingdom), the National Research Council (Canada), CONICYT (Chile), the Australian Research Council (Australia), MinistŽrio da Cincia, Tecnologia e Inova‹o (Brazil) and Ministerio de Ciencia, Tecnolog'a e Innovaci—n Productiva (Argentina).
CL is the recipient of an Australian Research Council Future Fellowship (project number FT0992259). This research was conducted by the Australian Research Council Centre of Excellence for All Sky Astrophysics (project number CE11E0900). 
MS acknowledges support from the Royal Society. The Liverpool Telescope is operated on the island of La Palma by Liverpool John Moores University in the Spanish Observatorio del Roque de los Muchachos of the Instituto de Astrofisica de Canarias with financial support from the UK Science and Technology Facilities Council. This publication has been made possible by the participation of more than 10\,000 volunteers in the Galaxy Zoo Supernovae project, http://supernova.galaxyzoo.org/authors.
A. G. acknowledges support by ISF, BSF, GIF and Minerva grants, an ARCHES award, and the Lord Sieff of Brimpton Fund.
S.B.C. and acknowledges generous financial assistance from Gary \& Cynthia Bengier, the Richard \& Rhoda Goldman Fund, NASA/{\it Swift} grants NNX10AI21G and GO-7100028, the TABASGO Foundation, and NSF grant AST-0908886.
MMK acknowledges generous support from the Hubble Fellowship and Carnegie-Princeton Fellowship.
\vspace{-5mm}
\newcommand{\noopsort}[1]{}
\bibliographystyle{mn2e}
\bibliography{paper}
\label{lastpage}
\end{document}